\documentclass[reprint]{revtex4-2}
\usepackage[final]{graphicx}
\usepackage{dcolumn}
\usepackage{bm}
\usepackage{amsmath}
\usepackage{amssymb}
\usepackage{amsfonts}
\usepackage{fancyhdr}
\usepackage{slashed}
\usepackage{bbm}
\usepackage{latexsym,epsfig,bbm}
\usepackage[colorlinks=true,urlcolor=blue,citecolor=blue]{hyperref}
\usepackage[table,xcdraw]{xcolor}
\usepackage[braket]{qcircuit}

\usepackage[normalem]{ulem}

\newcommand{\figref}[1]{Fig.~\ref{#1}}
\newcommand{\tabref}[1]{Table~\ref{#1}}
\newcommand{\secref}[1]{Sec.~\ref{#1}}

\usepackage{color}
\definecolor{blue}{rgb}{0,0.2,1}

\definecolor{red}{rgb}{0.9,0,0}

\begin{document}

\title{Testing of flag-based fault-tolerance on IBM quantum devices}

\author{Anirudh Lanka$^e$} 
\email{alanka@usc.edu}

\affiliation{%
 Ming Hsieh Department of Electrical Engineering, University of Southern California, Los Angeles, California 90089, USA
}

\date{\today}

\begin{abstract}
It is hard to achieve the theoretical quantum advantage on NISQ devices. Besides the attempts to reduce the error using methods like error mitigation and dynamical decoupling, small quantum error correction and fault-tolerant schemes that reduce the high overhead of traditional schemes have been proposed as well. According to \cite{chao2018quantum, chao2018fault, chamberland2018flag}, it is possible to minimize the number of ancillary qubits using flags. While implementing those schemes is still impossible, it is worthwhile to bridge the gap between the NISQ era and the FTQC era. Here, we introduce a benchmarking method to test fault-tolerant quantum error correction with flags for the $[[5,1,3]]$ code on NISQ devices. Based on results obtained using IBM's \textit{qasm} simulator and its $15$-qubit \textit{Melbourne} processor, we show that this flagged scheme is testable on NISQ devices by checking how much the subspace of intermediate state overlaps with the expected state in the presence of noise.
\end{abstract}

\maketitle

\section{Introduction}

Quantum computation could provide computational speedups compared to their classical counterparts on some specific types of classically intractable problems. However, it is tricky to achieve this theoretical advantage on current quantum devices in this noisy intermediate-scale quantum (NISQ) era \cite{preskill2018quantum}. Quantum error correction (QEC) and fault-tolerant quantum computation (FTQC), having been studied for decades, could delicately handle and even tolerant the errors occurred during the computation procedure. Traditional QEC and FTQC schemes are resource-intensive since they require a large number of ancillary qubits to store the redundancy information for checking stabilizers. To reduce the high overhead, one significant progress made recently is adding flags to the original schemes \cite{chao2018quantum, chao2018fault, chamberland2018flag}. As proposed in \cite{chao2018fault}, this flag scheme allows us to implement QEC schemes on quantum computer with only two extra qubits. Still, it is beyond those noisy, shallow and small devices' ability to implement a whole QEC or FTQC scheme for codes more complex than the repetition code. While lots of studies at the moment focus on more near-term methods such as quantum error mitigation and dynamical decoupling, it is still worthwhile to bridge the gap between FTQC and the NISQ era. Here, we introduce a benchmarking method to test only one stabilizer of fault-tolerant quantum error correction with flags on NISQ devices, which is simple, scalable and flexible. We ran simulation on IBM's \textit{qasm} simulator and experiment on its $15$-qubit \textit{Melbourne} device. By post processing the data, we show that the reconstruction result of the final state can fit well to theoretical noise model.

We start by a brief introduction to fault-tolerant quantum error correction with flags in Sec. II. Our proposal is developed in Sec. III. The implementation on IBM's QPU is explained in Sec. IV. In Sec. V, we show our simulation and experimental results. A summary of our work and some open questions are listed in Sec. VI.

\section{Fault-tolerant quantum error correction with flags\label{sec:flagftqc}}
One of the fundamental building blocks of fault-tolerant quantum computing is the ability to perform syndrome measurement fault-tolerantly. Steane-style fault-tolerance for CSS codes requires as many ancilla qubits as data qubits\cite{steane}. Shor-style fault tolerance requires $w+1$\cite{Shor} or $w$\cite{DiVinAlif} ancilla qubits, where $w$ is the largest weight of a stabilizer generator of the code. For $[[2^r-1, 2^r-1-2r,3]]$ Hamming codes, one would require $2^{r-1}+1$ ancilla qubits. Realization of such a circuit on near-term devices is not possible due to the large number of qubits and circuit depth. To counter this issue, several algorithms have been proposed to reduce this number to $2^{r-2}$ (Decoded Half-cat)\cite{chao2018quantum}. But this is still of exponential order. Using the procedures outlined in \cite{chao2018quantum}, we can achieve fault-tolerant syndrome measurement on certain families of codes using only $2$ extra qubits.

\begin{figure}
    \centering
    \includegraphics[scale=.4]{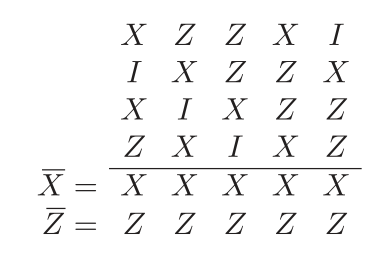}
    \caption{Stabilizer generators and logical operators of $[5,1,3]$ code\cite{chao2018quantum}}
    \label{fig:gen_log_ops}
\end{figure}
 
\begin{figure}
    \centering
    \includegraphics[scale=.4]{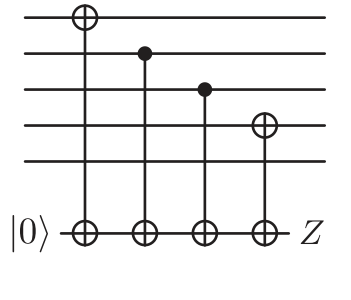}
    \caption{Circuit to extract the syndrome of the $XZZXI$ stabilizer non-fault-tolerantly.\cite{chao2018quantum}}
    \label{fig:non_ft}
\end{figure}

\begin{figure}
    \centering
    \includegraphics[scale=.4]{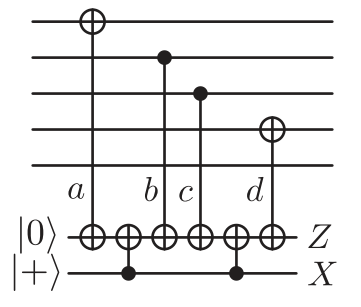}
    \caption{Circuit to extract the syndrome of the $XZZXI$ stabilizer fault-tolerantly. \cite{chao2018quantum}}
    \label{fig:flag_ft}
\end{figure}

Flagged error correction for the $[[5,1,3]]$ code – The logical operators and stabilizer generators of the perfect $[[5,1,3]]$ code are given in \figref{fig:gen_log_ops}. A circuit to measure the stabilizer $XZZXI$ non-fault-tolerantly is given in \figref{fig:non_ft}. It is not fault-tolerant since errors on the ancilla qubit can lead to weight-$2$ errors on the corresponding data qubits. A flagged fault-tolerant circuit to measure the stabilizer $XZZXI$ is shown in \figref{fig:flag_ft}. An $X$ error on the syndrome qubit will never spread to data qubits. However, $Y$ and $Z$ errors can spread and hence we use a flag qubit to catch them. Without any errors, this circuit behaves the same as that in \figref{fig:non_ft}, and measurement of the flag qubit in the $X$-basis will always give a $\ket +$ state. When there are faults in the syndrome qubit, the flag's $X$-basis measurement may result in the  $\ket -$ state indicating the presence of $Y$ or $Z$ errors in the syndrome qubit. The process is repeated for all stabilizer generators to extract all syndromes and flags. 

Since the codes we consider are distance-$3$ fault-tolerant, we assume that our error correction block must also be distance-3 fault-tolerant. Hence if we catch one fault occurring during syndrome extraction, we assume there are no further faults. If the syndrome extracted by a flagged circuit is nontrivial, even when the flag is not raised, this may be because of a data qubit fault in the middle of syndrome extraction. So, all the syndromes and flags are non-fault-tolerantly extracted before applying any correction procedures \cite{chao2018quantum}. This will ensure all the data qubits are free from any obvious errors.

\section{Proposal\label{sec:proposal}}

In order to estimate the efficiency of the flagged fault-tolerant quantum error correction scheme, we outline what an ideal experiment would look like.

\begin{itemize}
    \item First, we prepare a code state of the $[[5,1,3]]$ code fault-tolerantly. Using the circuit in \figref{fig:ftstateprep}, we can prepare the $\ket{-}_L$ state. This consists of the non-fault-tolerant preparation of the $\ket{-}_L$ state, which is also the $5$ cycle graph state, followed by the measurement of $3$ of its stabilizers. The stabilizers are measured using flagged syndrome qubits to achieve fault-tolerance.
    \item Once the data qubits are initialized with the encoded state, the circuit in \figref{fig:flag_ft} is used to measure the syndrome and flag qubits of the $XZZXI$ stabilizer. 
    \begin{itemize}
        \item If the flag qubit is measured as $\ket -$, then the circuit analogous to \figref{fig:non_ft} is used to extract all $4$ syndromes. The resulting weight $\geq 1$ data error can be identified by the non -fault-tolerant syndrome measurement.
        \item If the flag qubit is measured as $\ket +$, the remaining $3$ syndrome and flag qubits are measured using the circuits analogous to \figref{fig:non_ft} and \figref{fig:flag_ft} to determine the necessary correction operators. During the measurement of any of these stabilizers, if the flag is triggered, we revert to measuring all four stabilizers non-fault-tolerantly.
    \end{itemize} 
    \item State tomography is performed on the data qubits to determine its density matrix after applying the correction procedures.
    \item Finally, the fidelity is calculated between the state of the data qubits before and after the correction procedure.  
\end{itemize}

\begin{figure}
    \centering
    \includegraphics[scale=.4]{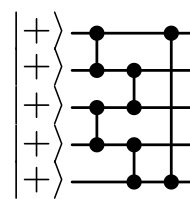}
    \caption{State preparation for the 5-cycle graph state, i.e., the encoded $\ket{-}_L$ state of the $[[5,1,3]]$ code. \cite{chao2018quantum}}
    \label{fig:graph_state}
\end{figure}

\subsection{Shortcomings of near-term devices}

Access to a quantum computer has never been easier. Current quantum cloud devices, such as those hosted by IBM and Rigetti, have paved the way for accelerated research in quantum computing. However, the technology lags behind the theory. In our pursuit to run flag-fault-tolerant circuits on an IBM device, we faced some challenges. Unfortunately the technology does not currently exist to reset individual qubits in either IBM's or Rigetti's quantum devices. Aside from this, measurement times on these devices are prohibitively large. In the time it takes to measure a qubit, other qubits that are not being measured will have decohered to noise. Hence, post-measurement operations (and thereby classically controlled operations) are not allowed on IBM's quantum devices as well.
Altogether, this rules out the possibility of requiring only $2$ extra qubits to measure all four stabilizers. Instead, we could measure all the stabilizers using unique pairs of syndrome and flag qubits, resulting in a total of $8$ ancillas. As discussed in \secref{sec:device}, a circuit to perform this would, however, be too large in depth and use too many noisy gates to produce meaningful results.

Due to the above constraints on measurement times and qubit reset times, we considered measurement-free error correction. In existing measurement-free schemes\cite{crow2016improved}, the ancillary qubits store only classical information and classically controlled operations are replaced by $C_k NOT$ gates, which can be decomposed into $2(k-1)$ Toffoli gates and $1$ CNOT gate. Only two more extra qubits are required to perform this, and by swapping qubits we can ideally implement on IBM's 15-qubit \textit{Melbourne} processor. 
However, Toffoli gates are unfavorable on topologies of current lattice-like quantum devices due to the extra ancilla and swap cost. Also, this replacement would make the circuit depth much larger than the original flag scheme, which could undermine any attempt of error correction. 

\subsection{Reformulation of goals}

Due to the challenges listed above, we were forced to reformulate our experiment.

\begin{itemize}
    \item Run non-fault-tolerant state preparation. 
    \item Measure one stabilizer with a flag. (All qubits are measured at this stage.)
    \item Based on the flag outcomes, post-process the data qubits as follows:
    \begin{itemize}
        \item f=0: Perform stabilizer measurement of the other three stabilizers. Keep all stabilizer measurement results.
        \item f=1: Throw away result of current stabilizer measurement. Perform stabilizer measurement of all four stabilizers. 
    \end{itemize}
    \item Use stabilizer measurement results to virtually apply the necessary correction.
    \item Verify that the data qubits have returned to the $+1$ eigenspace of the code.
\end{itemize}

The target experiment described above is the smallest experiment one can perform to measure a stabilizer using flagged fault-tolerance.

\section{Implementation on the IBM QPU}

\subsection{Device}
\label{sec:device}
To test the efficiency of flagged fault-tolerant schemes with the $[[5,1,3]]$ code (which has 4 stabilizer generators), we need to extract $4$ pairs of syndromes and flags. But we face the following issues with the proposed experiment: 

\begin{figure}
    \centering
    \includegraphics[scale=.4]{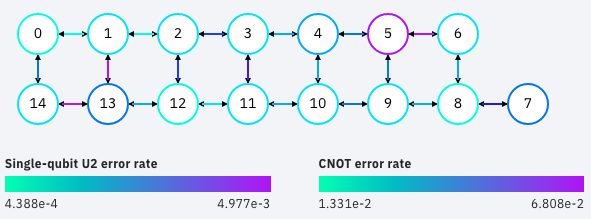}
    \caption{IBMQ \textit{Melbourne} Layout and Error Rates \cite{Qiskit}}
    \label{fig:ibm_device}
\end{figure}

\begin{itemize}
    \item Current IBM quantum computers do not support RESET operations. Hence, each stabilizer measurement must be performed on a different ancilla qubit. 
    \item The circuit used for measuring all the 4 stabilizers will produce results that fall well below the accepted limit of infidelity, due to the large depth, and relatively high error rates 
\end{itemize}

Considering the above factors, and the fact that we require 7 qubits in total, the IBM \textit{Melbourne} backend with 15 operational qubits is chosen for running the experiments. The gate layout with single-qubit U2 error rates and CNOT error rates is given in the \figref{fig:ibm_device}. We particularly notice that the qubits $5$ and $13$ have high U2 error rates and the connections $(1,13),(5,6),(13,14)$ have high CNOT error rates, suggesting minimal usage of those gates. 

The native gates available with this device are listed in \tabref{tab:melbourne_gates} along with the associated gate times and error rates.

\begin{table}
    \centering
    \begin{tabular}{|c|c|c|}
         \hline 
         Gate & Gate duration ($\mu$s) & Gate error (\%) \\
         \hline 
         id & $ 0.0978 \pm 0.0444$ & $-$\footnote{'id' gate errors are due to the environment and not due to imperfect pulses.} \\
         u1 & $0$ & $0$ \\
         u2 & $ 0.0978 \pm 0.0444$ & $0.27 \pm 0.227$ \\
         u3 & $ 0.1955 \pm 0.0889$ & $0.54 \pm 0.452$ \\
         cnot & $ 1.002 \pm 0.647$ & $4.069 \pm 2.739$ \\
         \hline 
         \hline 
    \end{tabular}
    \caption{Gate duration and error rate ranges calculated based on properties extracted from the device backend. These values change daily, as and when IBM calibrates the device.}
    \label{tab:melbourne_gates}
\end{table}

\subsection{Preprocessing - Circuit transpilation}

\begin{figure}
    \centering
    \includegraphics[scale=.25]{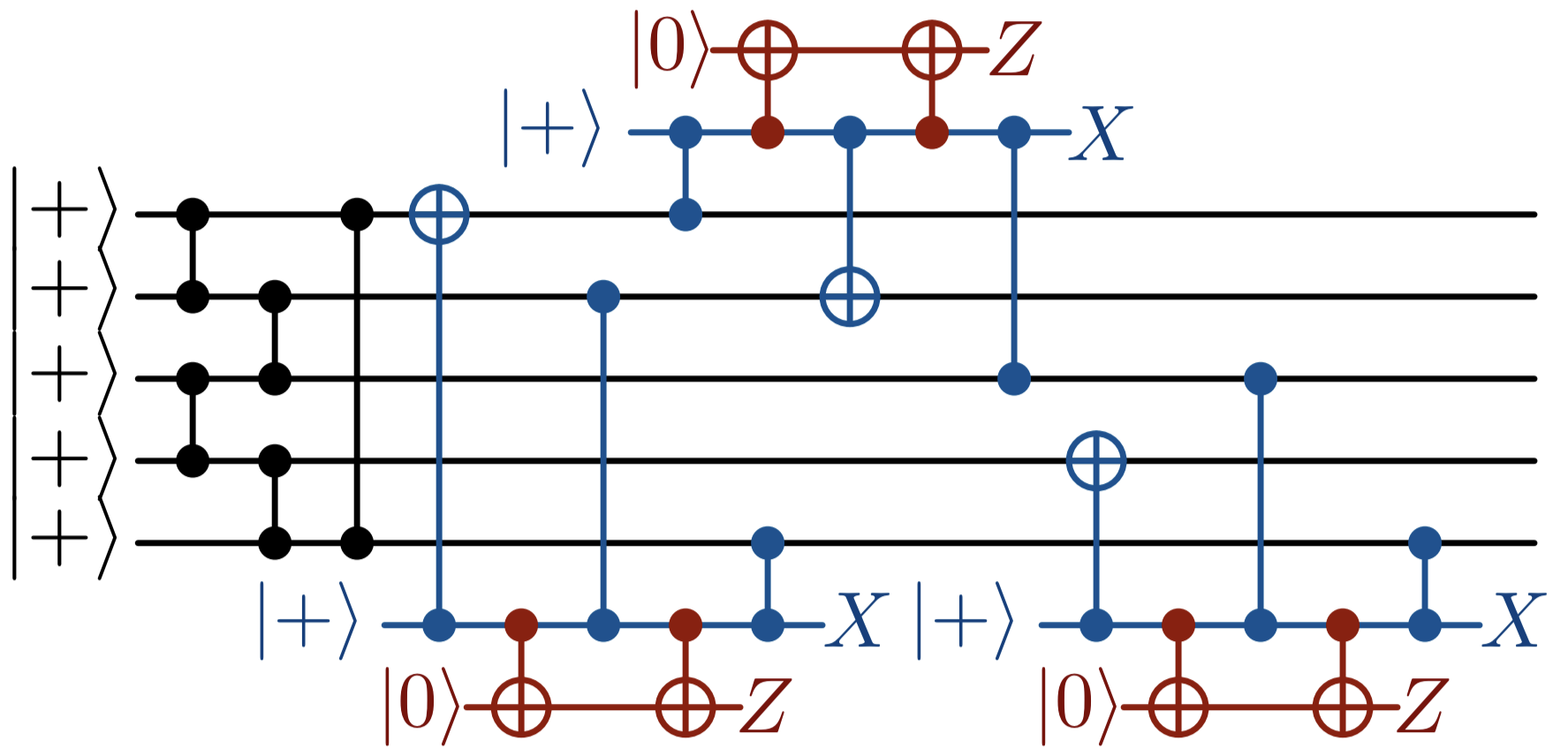}
    \caption{Fault-tolerant $\ket{-}_L$ state preparaton requires the measurement of at least three stabilizers for verification. The stabilizer measurement is done using flag-based fault-tolerance\cite{chao2018quantum}.}
    \label{fig:ftstateprep}
\end{figure}

Fault-tolerant state preparation using flags involves the measurement of multiple stabilizers to verify the correct state has been prepared after non-fault-tolerant state preparation. This is shown in \figref{fig:ftstateprep}. Due to constraints on the depth of circuits run on the QPU, we chose to perform stabilizer measurement on a non-fault-tolerantly prepared $\ket{-}_L$ state. 

In order to prepare the $\ket{-}_L$ state of the $[[5,1,3]]$ code, it suffices to prepare the five cycle graph state. Without constraints on the connectivity of the qubits, it is possible to implement this circuit in depth $3$ as shown in \figref{fig:graph_state}. 
While implementing this circuit on the 15 qubit \textit{Melbourne} QPU, we require transpilers to convert ideal circuits to circuits that respect connectivity constraints. The first choice was to use IBM's in-house transpiler. However, we quickly identified that it uses a stochastic algorithm and would return different optimizations each time the transpiler was run.

\begin{figure}
    \centering
    \includegraphics[scale = 0.8]{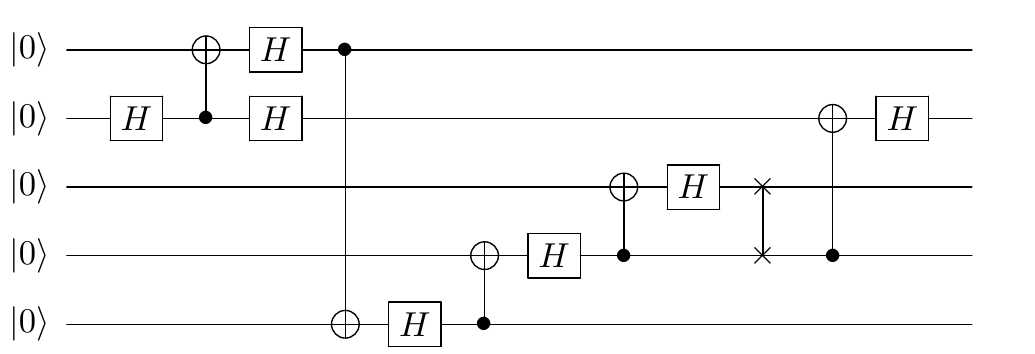}
    \caption{Circuit transpiled by pytket that respects the connectivity constrainsts of the IBM \textit{Melbourne} device.}
    \label{fig:pytketstateprep}
\end{figure}

Our next option was to use pytket\cite{tket}, a python package with tools to deterministically route circuits while also reducing the number of noisy gate operations. Although pytket's transpiled circuit, \figref{fig:pytketstateprep}, was consistent over multiple runs, we felt further optimizations could be made.

\begin{figure}
    \centering
    \includegraphics[scale = 1.0]{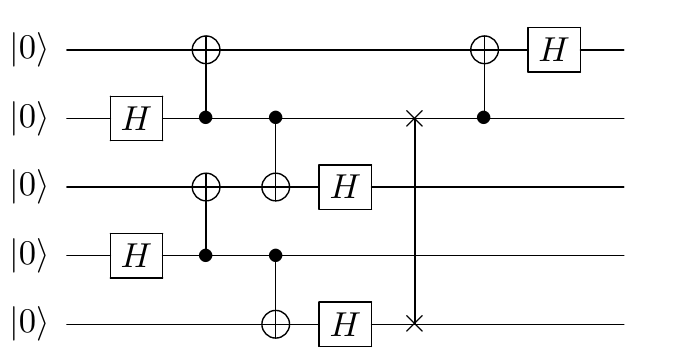}
    \caption{Manually optimized circuit. Starting initial layout on the device was also chosen more optimally than previous transpilers}
    \label{fig:manualstateprep}
\end{figure}

\begin{figure}
    \centering
    \includegraphics[scale = 1.0]{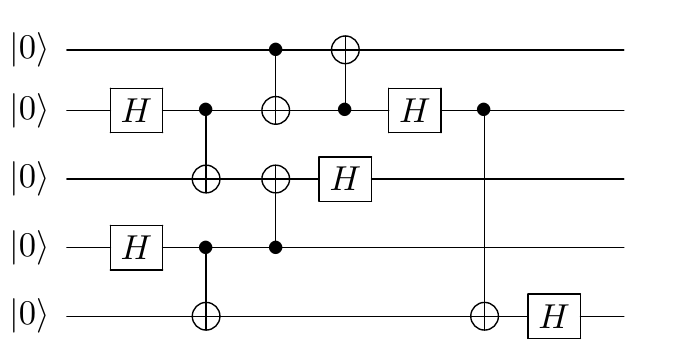}
    \caption{A circuit to prepare $\ket{-}_L$ of the $[[5,1,3]]$ code with CNOT depth $= 4$}
    \label{fig:manualstateprep2}
\end{figure}

By manually optimizing the circuits produced by pytket, we were able to define two circuits, one of depth $6$, \figref{fig:manualstateprep}, and another of depth $4$, \figref{fig:manualstateprep2}.

\begin{figure}
    \centering
    \includegraphics[scale = 0.7]{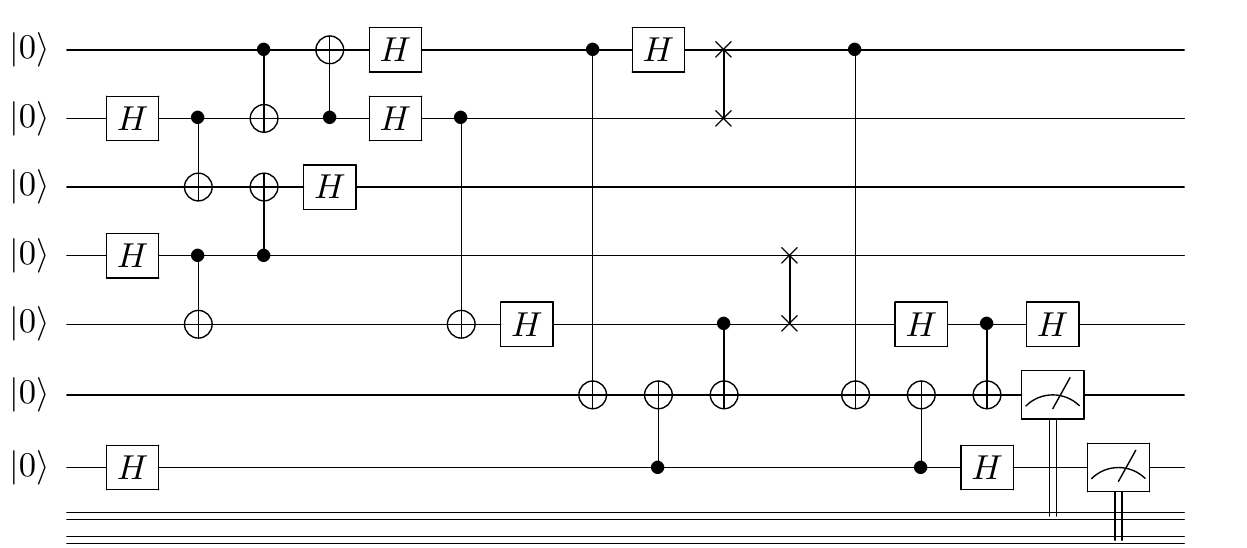}
    \caption{$ZXIXZ$ stabilizer measurement circuit with CNOT depth $= 10$. Qubits $q[0]$ through $q[4]$ are the data qubits. $q[5]$ and $q[6]$ are the syndrome and flag qubit respectively.}
    \label{fig:manualstabmeas}
\end{figure}

Using the depth-$4$ circuit also proved useful in determining a low-depth stabilizer measurement circuit. Using two additional qubits and with just two swap operations, we are able to measure the $ZXIXZ$ stabilizer with a flag, as shown in \figref{fig:manualstabmeas}. The stabilizer order is with respect to the initialized physical qubits, $q[0]$ to $q[4]$. The overall CNOT depth of the circuit is $10$, although there are $18$ CNOT gates in total.

\subsection{Postprocessing}
\subsubsection{Conditional quantum state tomography}

While collecting results for the experiments defined above, it is not enough to only measure in the $Z$ basis. Since we want to post-process the state of the data qubits, we require the full information of the state. We were presented with two options to collect this information. We could either perform state tomography, which would require an exponentially large number of runs (and post-processing to boot), or we could run the circuit in reverse and measure in the $Z$ basis. The shortcoming of the second method was illustrated by running state tomography on the state preparation circuit. Short circuits of depth $4$-$6$ created the $\ket{-}_L$ state only with $35\%$ fidelity. Running the circuit in reverse would, in this case, hurt the fidelity more than it would save us time.

Since our objective involved executing a circuit with 7 qubits, we naturally assumed all the qubits would need to be tomographed, resulting in $3^7 = 2187$ circuits. However, upon closer inspection, we noticed only the 5 data qubits needed to be tomographed, and that the remaining two qubits are always measured in the computational basis. This reduced the number of measurement bases to $3^5 = 243$.

IBM's qiskit-ignis \cite{Qiskit} provides functions to perform conditional and unconditional state tomography. The density matrix reconstruction algorithm goes beyond naively performing direct inversion. It provides the option of using either a least-squares algorithm or convex optimization algorithm to find a valid density matrix. 

\subsubsection{Density matrix reconstruction and error post-processing\label{subsubsec:dm}}

We reconstruct the density matrix, w.l.o.g., and denote it as $\tilde{\rho}$. Ideally we would expect the syndrome and flag outcomes to always be zero and the remaining five-qubit system $Tr_{anc}(\tilde{\rho})$ is close to the ideal density matrix $\rho=|-\rangle\langle -|$. However, there are several reasons that the reconstructed density matrix doesn't satisfy this: 
\begin{enumerate}
    \item The finite number of shots could introduce statistical error.
    \item The strong decoherence and noise of the system may drive down the fidelity.
    \item For the flagged circuit, fault tolerance itself allows weight-one errors at the end of any step of the procedure, hence even if we perform the correction, the resulting codespace will be similar to the one passed through a depolarizing channel.
\end{enumerate} 
The third point above is not as intuitive as the first two, so we explain why it matters based on the circuit in Fig. \ref{fig:manualstabmeas}:

Since we use error mitigation methods to process the tomography results, here we only consider that the error occurs after gates in stabilizer measurement for simplicity. Assume there is only one error $P_2 Z_4$, where the indices $2$ and $4$ represent the location at the end, right after the first syndrome extraction gate, where $P_2$ is uniformly and randomly chosen from $\{I_2, X_2, Y_2, Z_2\}$. Then the data error would be $P_2$ and the measurement outcomes are both zero. Assume the probability of an error occuring is $p$ in a one-qubit depolarizing channel
\begin{equation*}
\rho \rightarrow p\rho + \frac{1-p}{3}(X\rho X + Y\rho Y + Z\rho Z) = p\rho + \frac{2(1-p)}{3} I.
\end{equation*}
Also note that a failure after the fourth stabilizer gate can also lead to all-zero measurement outcomes on the syndrome and flag. Therefore, the procedure would be like to pass the $2$nd and the $4$th qubits of the ideal one through such a depolarizing channel independently with the probability of $p$. When we run multiple rounds of experiment and strip out all outcomes except $00$, the density matrix of the data qubits $\tilde\rho_{data}=\sum_i \rho_{135, i} \otimes \rho_{2, i} \otimes \rho_{4, i}$ on average will evolve as:
\begin{equation}
\begin{aligned}
    \sum_i & \rho_{135, i} \otimes \rho_{2, i} \otimes \rho_{4, i} \\
    \rightarrow \sum_i & [ p^2 \rho_{135, i} \otimes \rho_{2, i} \otimes \rho_{4, i} + (\frac{2(1-p)}{3})^2 (\rho_{135, i} \otimes I_4 \otimes I_4)\\
    + & \frac{2p(1-p)}{3} (\rho_{135, i} \otimes I_2 \otimes \rho_{4,i} + \rho_{135, i} \otimes \rho_{2,i} \otimes I_4)], 
\end{aligned}\label{eq:pchannel}
\end{equation}
which could help us find the gate error rate and verify the code space after the stabilizer. 

\subsubsection{Measurement error mitigation}

IBM's qiskit-ignis \cite{Qiskit}  package also provides tools to mitigate measurement errors. The general idea is that we can construct a measurement calibration matrix for any device, using two properties of measurement operations: a) the probability of measuring 0 when in state $\ket 1$, and b) the probability of measuring 1 when in state $\ket 0$.

We first performed a simulation of the state preparation circuits with a noise model constructed using the measurement error properties listed above. Since the experiment is performed on a simulator, one would ideally expect $\simeq 99\%$ fidelity. We used a noise model corresponding to the measurement error rates of the 15 qubit \textit{Melbourne} device, and achieved only $71\%$. Upon recalibrating our results using the calibration matrix, we were able to boost this fidelity back up to $99\%$. Since our noise model was derived from the \textit{Melbourne} device, the resulting calibration matrix could be used on all further experiments (for the calibrated parameters of any given day). 

\section{Results}

Using the code provided at \cite{githubpage}, we obtained the following results.

We ran the circuit for $[[5,1,3]]$ code state preparation (\figref{fig:manualstateprep2}) and the circuit extended by the $ZXIXZ$ stabilizer (\figref{fig:manualstabmeas}) on IBM's \textit{Melbourne} processor. As shown in \tabref{tab:resdetail}, extra running time $\simeq 4.32\mu s$ is introduced by the stabilizer measurement. Thereby, the increasing circuit depth leads to strong decoherence despite the fact that the total time $\simeq 7.58\mu s$ looks acceptable given the energy relaxation time $T_1$ and the dephasing time $T_2$ are both $\geq 50 \mu s$. More specifically, to show the effect of decoherence more directly, we executed a third circuit: state preparation extended by $106$ repeating cycles consisting of one identity gate and one barrier on the first qubit, whose runtime is close to state preparation followed by $ZXIXZ$ stabilizer measurement.

The third column of \tabref{tab:resdetail}(a) shows the fidelity between the reconstructed states and the expected final state simulated by \textit{qasm} in the absence of noise. For the two circuits with large depth, state preparation with identity gates and state preparation with one stabilizer, the fidelity is unfortunately below $5\%$ due to decoherence; For the circuit including only state preparation, the fidelity could achieve at least $\simeq 40\%$.

\begin{figure}
    \centering
    \includegraphics[scale = 0.8]{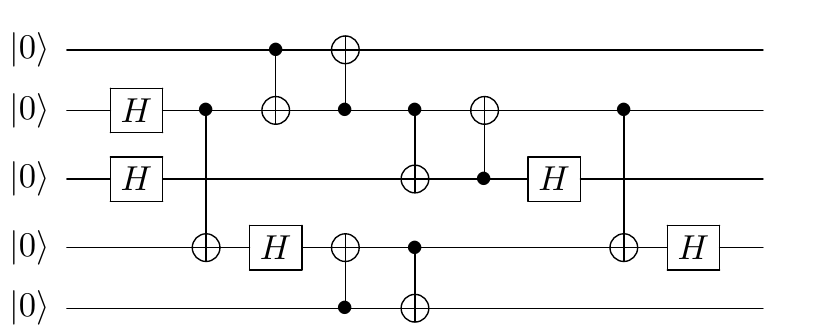}
    \caption{Manually transpiled circuit for $\ket{-}_L$ state preparation using the IBM \textit{Vigo} device. The order of the physical qubits of the code at the input and output are (1,2,0,3,4) and (2,0,1,4,3)}
    \label{fig:vigostateprep}
\end{figure}

We show that, while state preparation on the \textit{Melbourne} device only yields a maximum of $40.69\pm 2.288\%$ fidelity, we can use other devices to achieve better fidelity. As an example, we ran a manually transpiled circuit, \figref{fig:vigostateprep} on the $5$ qubit IBM \textit{Vigo} machine, and over three runs, we were rewarded with a state fidelity of $81.32\pm 2.216 \%$

\begin{table}[h]
    \centering
    \begin{tabular}{|c|c|c|}
    \hline
        (a) Circuit & Runtime($\mu s$)\footnote{Runtime is defined as the maximum time for each qubit to interact with gates before measurement} & Fidelity\\
        \hline
        State prep. (A) &  $2.905$ & $38.32\pm 5.486\%$ \\ 
        State prep. (B) &  $3.264$ & $40.69\pm 2.288\%$ \\
        State prep. (A) + $106$ I gates&  $7.460$ & $4.515\pm 1.163\%$\\
        State prep. (B) + $106$ I gates&  $7.872$ & $3.542\pm 0.295\%$\\
        State prep. (C) + $ZXIXZ$ & $7.584$ & $2.718\pm 0.254\%$\\  
        \hline
        \hline
        (b) Backend Properties & \multicolumn{2}{c|}{Value}\\
        \hline
        backend\_name & \multicolumn{2}{c|}{ibmq\_16\_melbourne} \\
        backend\_version & \multicolumn{2}{c|}{$2.1.0$} \\
        last\_update\_date & \multicolumn{2}{c|}{2020-05-11 09:26:40} \\
        gate length of $I_0$ & \multicolumn{2}{c|}{$53.3ns$}\\
        averaged gate length of $cx$ & \multicolumn{2}{c|}{$904.56ns$} \\
        minimum $T_1$ in C ($T_{1_{q_{13}}}$) & \multicolumn{2}{c|}{$24.785\mu s$} \\
        minimum $T_2$ in C ($T_{1_{q_4}}$) & \multicolumn{2}{c|}{$30.781\mu s$} \\
        \hline
        backend\_name & \multicolumn{2}{c|}{ibmq\_vigo} \\
        backend\_version & \multicolumn{2}{c|}{$1.0.2$} \\
        last\_update\_date & \multicolumn{2}{c|}{2020-05-12 07:02:30} \\
        gate length of $I_0$ & \multicolumn{2}{c|}{$35.5ns$}\\
        averaged gate length of $cx$ & \multicolumn{2}{c|}{$344.7ns$} \\
        maximum $T_1$ in C ($T_{1_{q1}}$) & \multicolumn{2}{c|}{$23.09\mu s$} \\
        maximum $T_2$ in C ($T_{1_{q1}}$) & \multicolumn{2}{c|}{$15.4\mu s$} \\
        \hline
    \end{tabular}
    \caption{Implementation details: (a) circuits and performance, and (b) backend properties. A, B and C represent different initial layouts: A=(1, 2, 3, 11, 12); B=(11, 12, 13, 1, 2); C=(11, 12, 13, 1, 2, 3, 4). State preparation with A will produce the best fidelity, but we need B to compare against the case when we perform the $ZXIXZ$ stabilizer.
    \label{tab:resdetail}}
\end{table}

However, the fidelity only explains the decoherence. To prove or disprove the advantage of FTQC on NISQ devices, we need to carefully study the influence of stabilizer on the change of code space as mentioned in \secref{subsubsec:dm}. Here we simply focus on the case when the syndrome and the flag are both measured as $0$. We do not need to worry about correlated error and the error at the end could be $P_1$ or $P_4$ ($P\in \{I, X, Y, Z\}$, either introduced by a fault after the $1$th gate or the $4$th in stabilizer measurement. Then using \eqref{eq:pchannel}, we could find the optimum $p$, the probability of getting no error at the end, that minimizes the spectral norm between the reconstruction result and the resulting density matrix by passing an ideal 5-cycle graph state through a depolarizing channel parameterized by $p$. Specifically, for the result of state preparation (C) + $ZXIXZ$ in \tabref{tab:resdetail}, we have $p_{OPT}=0.64258$, which implies that the error rate on the first and the last data qubit is $\simeq 11.9\%$. 

\section{Conclusions}

In this report, we provide a benchmarking technique for running fault-tolerant quantum error correction with flags on NISQ devices. Though the results are not so impressive, our verification method has some advantages:
\begin{itemize}
    \item Simplicity: Until gate error rates become low enough, we can test flag based fault-tolerance on quantum devices by measuring just one stabilizer fault-tolerantly. Post-processing of the resulting density matrix can allow us to identify errors to some extent.
    \item Flexibility: Since we measure only one stabilizer of the code on the quantum processor, there is flexibility in the choice of stabilizer. 
    \item Scalability: Since the stabilizers are independent of each other, measuring them in parallel could have lower logical error rates\cite{flagsAndBridges}. As long as we have access to a larger quantum computer with decent error rates for the two-qubit entangling gate, our method could be easily generalized to test more about FT state preparation and measurement.
    
\end{itemize}

Our project suggests some further extensions, as well as some interesting open problems.
\begin{itemize}
    \item In \secref{subsubsec:dm}, we only consider the effect of independent gate faults in the syndrome extraction process. However, in the real world, faults can occur everywhere and unentangled qubits could influence others, which would necessitate a more complex noise model to analyze the reconstructed final density matrix.
    \item The topology and entangling-gate error rates on IBM's open-access devices are some of the biggest bottlenecks that restrict the performance of our benchmarking method. Given devices with better qubit connectivity and more robust gates, we can measure more than just one stabilizer. Given the right topology, we could also measure all four stabilizers in parallel.
    \item With qubit measurement times on a downward trend, it seems inevitable that one day we will be able to measure syndromes and apply classically controlled quantum operations. Hints of this technology are appearing in Honeywell devices \cite{pino2020demonstration}, so we predict that the ideal proposal in \secref{sec:proposal} can soon be implemented.
\end{itemize}

\section{Acknowledgements}

The authors would like to thank Daniel Lidar for teaching quantum error correction course and suggesting the study of this problem. We would also like to thank Bibek Pokharel for helpful and inspiring conversations. 

\nocite{*}

\end{document}